\documentclass[11pt]{article}
\usepackage{moriond,epsfig}

\def\Journal#1#2#3#4{{#1} {\bf #2}, #3 (#4)}

\def\apj{{\em Astrophys. J.}}
\def\aap{{\em Astron. Astrophys.}}
\def\asr{{\em Adv. Space Res.}}
\def\astl{{\em Astronomy Lett.}}
\def\nat{{\em Nature}}

\def\be{\begin{equation}}
\def\ee{\end{equation}}
\def\bea{\begin{eqnarray}}
\def\eea{\end{eqnarray}}

\def\etal{{\it et~al.}}

\def\Granat{\hbox{\it GRANAT}}
\def\4u{4U1724-307}
\def\SLX{\mbox{SLX1744-299/300}}
\def\fdg{\hbox{$.\!\!^\circ$}}

\def\arcmin{$^{\prime\,}$}
\def\deg{$^\circ$}
\def\ergs{erg s$^-1$}
\begin{document}
\centerline{\it To be published in the proceeding of the XXIInd Moriond astrophysics meeting ``The Gamma-Ray Universe''}
\vspace*{4cm}
\title{ REVIEW OF X-RAY BURSTERS IN THE GALACTIC CENTER REGION}

\author{A.A.LUTOVINOV, S.V.MOLKOV, S.A.GREBENEV, M.N.PAVLINSKY }

\address{High Energy Astrophysics Department, Space Reserach Institute, 
Profsoyuznaya 84/32, 117810 Moscow, Russia}

\maketitle\abstracts{Results of observations of X-ray bursters in 
the Galactic Center region carried out with the RXTE observatory and
the ART-P telescope on board GRANAT are presented. Eight X-ray
bursters (A1742-294, SLX1744-299/300, GX3+1, GX354-0, SLX1732-304,
4U1724-307, KS1731-260) were revealed in this region during five
series of observations which were performed with the ART-P telescope
in 1990-1992 and more than 100 type I X-ray bursts from these sources
were observed. For each of the sources we investigated in detail the
recurrence times between bursts, the bursts time profiles and their
dependence on the bursts flux, the spectral evolution of sources
emission in the persistent state and during the bursts. Two bursters
(SLX1732-304, 4U1724-307) located in the globular clusters Terzan 1
and 2, were investigated using the RXTE data as well. }

\section{Introduction}

Because of the high concentration of sources of different nature, the
Galactic-center region is one of the most interesting and most
commonly observed regions in X-rays. The first detailed X-ray map of
the Galactic center field was obtained by {\it EINSTEIN} in the soft
energy band ($\le4.5$~keV). In the harder band (4-30 keV) this region
was intensively studied with the \mbox{ART-P} telescope onboard
\Granat\ observatory (Pavlinsky \etal~\cite{pa1}). The total 
ART-P/\Granat\ exposure time of the Galactic center field observations
was $\sim$830 ks. Such a long exposure allowed to be investigated the
detailed X-ray map of the Galactic center region and to be obtained
the emission from persistent sources. Recently this region was
observed by the telescopes of the BeppoSAX (Sidoli {\em et
al.}~\cite{si}) and RXTE observatories.

In this paper we present results of observations of several X-ray
bursters located near the Galactic center carried out with the
ART-P/\Granat\ telescope in 1990-1992. These objects are regular
sources of type I bursts, which are thought to be resulted from
thermonuclear flares on the surface of accreting neutron stars.
Results of RXTE observations of two X-ray bursters SLX1732-304 and
4U1724-307, located in globular clusters Terzan 1 and Terzan 2,
respectively, are discussed too. \\

\section{Observations}
 
ART-P is a coded-mask telescope which is able to detect photons in the
3-60 keV energy band with the time resolution of 3.9~ms. The geometric
area of the position-sensitive detector is 625~cm$^{2}$. The telescope
field of view is 3\fdg4$\times$3\fdg6, the energy resolution is
$\sim$25\% at 5.9 keV and the instrument dead time is about 580 $\mu$s
(Sunyaev \etal~\cite{su}). The data transfer from buffer to the main
satellite memory happens each 150-200 sec and takes $\sim$30 sec. This
results in gaps between individual exposures in ART-P data sets.

The PCA instrument onboard the RXTE (Rossi X-ray Timing Explorer)
orbiting X-ray observatory consists of five identical proportional
counters with a total area of 6500 cm$^2$, the operating energy range
2-60 keV, and an energy resolution $<18$\% at 6 keV. Because of its
large area, the instrument is sensitive enough for a spectral analysis
of emission even from weak X-ray sources to be performed. The PCA
field of view is restricted by a collimator with a FWHM of
1\deg. Depending on peculiarities of the suggested study,
observational data during their initial onboard reduction can be
written in various telemetric formats. \\

\section{Results}

\begin{figure}[t]
\hspace{1cm}\psfig{figure=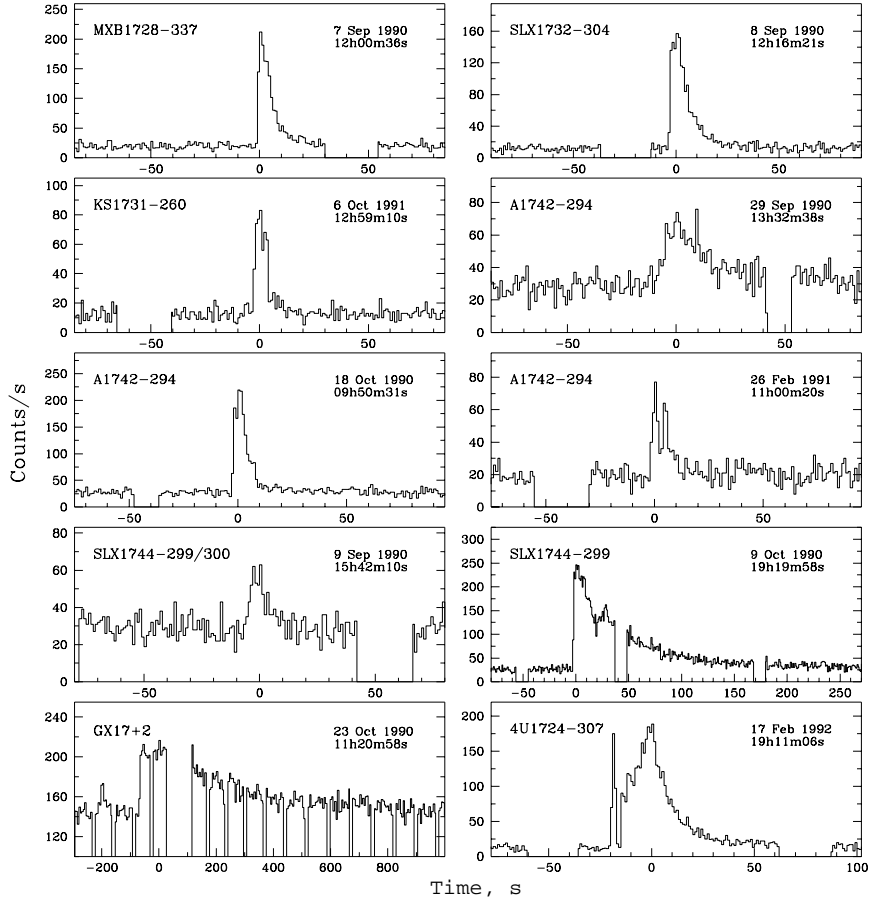,height=5in}
\caption{Examples of X-ray burst profiles, as observed with ART-P/\Granat\ 
in the 3-20 keV energy band.
\label{fig:obzor}}
\end{figure}

As was mentioned above X-ray bursts originat from the neutron stars
with weak magnetic field located in low-mass X-ray binaries. During
the ART-P/\Granat\ operation in 1990-1992 35 such systems were in the
field of view of the telescope and 112 X-ray bursts were detected from
11 from them. These bursts show a large variety in profiles, the most
interesting from them are shown in Fig.\,\ref{fig:obzor}.

\subsection{A1742-294} 

\begin{figure}[t]
\epsfxsize=130mm
\hspace{1cm}\epsffile[35 455 530 685]{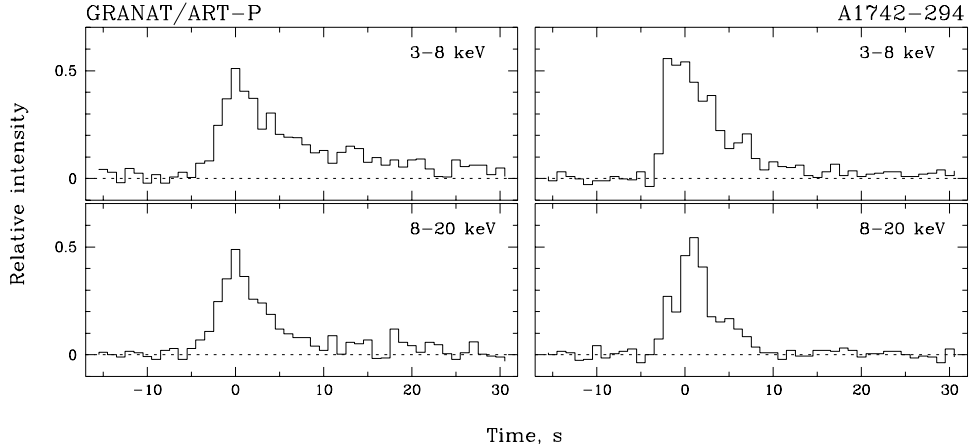}
\caption{The average profiles of ordinary and strong bursts detected from 
A1742-294 in two energy bands.
\label{fig:a17_avbur}}
\end{figure}

A1742-294 is the brightest persistent X-ray source in the Galactic
center field in the standard X-ray energy band. It is responsible for
$\sim1/3$ of the total X-ray emission from this region. The source
persistent flux in the 3-20 keV energy band varied in the range 20--50
mGrab during ART-P observations. This flux corresponds to a luminosity
of $(0.4-1)\times10^{37}$ \ergs~ assuming a distance of $\sim$8.5 kpc
(Pavlinsky \etal~\cite{pa1}). In total, 26 type I X-ray bursts with a
typical duration of $\sim(15-20)$~sec were identified with A1742-294
(Fig.\,\ref{fig:obzor}). Observations of several bursts from this
source during one session allowed us to measure the burst recurrence
time $t_r\sim$2.4~hr.

We found that the time profiles of the X-ray bursts detected from
A1742-294 depend on the source flux during the burst (Lutovinov
\etal~\cite{lu}). The shape of the burst profile is almost
``triangle'', i.e., the rise time of the burst is close to its decay
time, when the mean flux in the 3-20 keV energy was
$\sim100-300$~mCrab, whereas the burst profile becomes classical,
i.e., it is characterized by a sharp rise followed by a smooth decay,
when the flux increases to $\sim600-1000$~mCrab. To study this
dependence in detail, we constructed average profiles of ordinary and
strong bursts. We rejected bursts which occurred in the beginning or
the end of exposures; other bursts were normalized to the same peak
intensity and aligned by maxima of their intensity. In
Fig.\,\ref{fig:a17_avbur}, we show the average profiles of weak and
strong bursts in three energy bands. It is evident that the shapes of
weak and strong bursts in the hard energy band (8-20~keV) are similar,
while in the soft energy band (3-8 keV), the rise time of weak burst
is longer than that for strong bursts.

The mean energy release during the burst did not depend on the
source's pre-burst luminosity and was $E\simeq9\times10^{38}$ ergs for
ordinary bursts and $E\simeq2\times10^{39}$ ergs for strong bursts. To
provide such energy yields, $5.0\times10^{20}$ g and
$1.1\times10^{21}$ g of matter, respectively, must be accreted onto
the neutron-star surface. Assuming that the burster's mean luminosity
in quiescence is $L_{\rm X}\simeq8\times10^{36}$~\ergs, we can
estimate the characteristic times $\tau$ in which the required amount
of matter will be accreted onto the neutron-star surface:
$\tau\simeq3.2$ and $7.2$ h for ordinary and strong bursts,
respectively.

\subsection{SLX1744-299/300}

The other source located $\sim$1\deg\ from the GC and for which ART-P
detected several X-ray bursts was \SLX\ (Fig.\,\ref{fig:obzor}). This
source was discovered by {\it Spacelab-2} (Skinner \etal~\cite{sk})
and {\it Spartan-1} (Kawai \etal~\cite{ka}). In fact it is a double
X-ray source, with components SLX1744-299 and SLX1744-300 separated by
$\sim$3\arcmin, of which the latter is a burster, but the coordinates
of the extremely strong X-ray burst which was detected by ART-P from
this region coincide with the SLX1744-299 position, while SLX1744-300
was outside the position error cycle (Pavlinsky \etal~\cite{pa1}). The
average flux measured by ART-P from this double source was equal to
$\sim$20 mCrab.

\subsection{GX3+1}

This source was observed with ART-P four times during the \Granat\
Galactic center field survey in the fall of 1990 and the total
exposure exceeded 60 ksec. The quick-look analysis of the data led to
one interesting finding -- a strong X-ray burst was detected from
GX3+1 on Oct.\,14 (Fig.\,\ref{fig:obzor}). The burst duration
noticeably changed with energy being of about 12 s in the hard 10-20
keV band and reaching $\sim18$ s in the soft 2.5-6 keV band, what is
typically for the type I X-ray bursts (Molkov \etal~\cite{mo1}).
\begin{figure}[t]
\hbox{
\hspace{1cm}\begin{minipage}{5cm}
\hspace{0cm}\psfig{figure=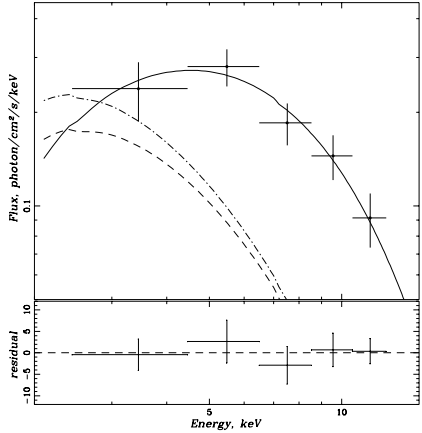,height=3in}
\end{minipage}
\hspace{2.5cm}\begin{minipage}{4.5cm}
\caption{The GX3+1 photon spectrum obtained with ART-P during X-ray 
burst on Oct. 14 and its best-fit approximation by the blackbody 
model (solid line). The dashed line present the persistent spectrum 
of the source measured on the same session, the dash-dotted line 
describe the spectrum observed in normal (non-bursting) state.
\label{fig:gx3_spbr}}
\end{minipage}}
\end{figure}
The source persistent luminosity $L_{p}$ was equal to
$(5.4\pm0.2)\times10^{37}\ \mbox{erg s}^{-1}$ in the 2.5-20 keV band
(assuming a distance of 8.5 kpc). This is $\sim 30$\% less than the
luminosity measured in the other days. The luminosity during the burst
interval was $L_{b}\simeq3\times10^{38}\, \mbox{erg s}^{-1}.$ The
black-body approximation of the burst average spectrum gave the
neutron star radius $R\simeq 7.2\pm1.2$ km and its surface temperature
$kT_b\simeq2.4\pm0.2$ keV. The source spectrum measured during the
burst is shown in Fig.\,\ref{fig:gx3_spbr} in comparison with spectra
of the source persistent emission observed in different states.

It is interesting to note that several bursts which were detected from
GX3+1 by HAKUCHO in 1980 (Makishima \etal~\cite{ma}) also occured then
the source was in the low X-ray state with the luminosity 30--50\%
smaller than the normal one.

\subsection{SLX1732-304}

\begin{figure}[t]
\hbox{
\begin{minipage}{70mm}
\epsfxsize=70mm
\hspace{-10mm}\epsffile[60 295 420 690]{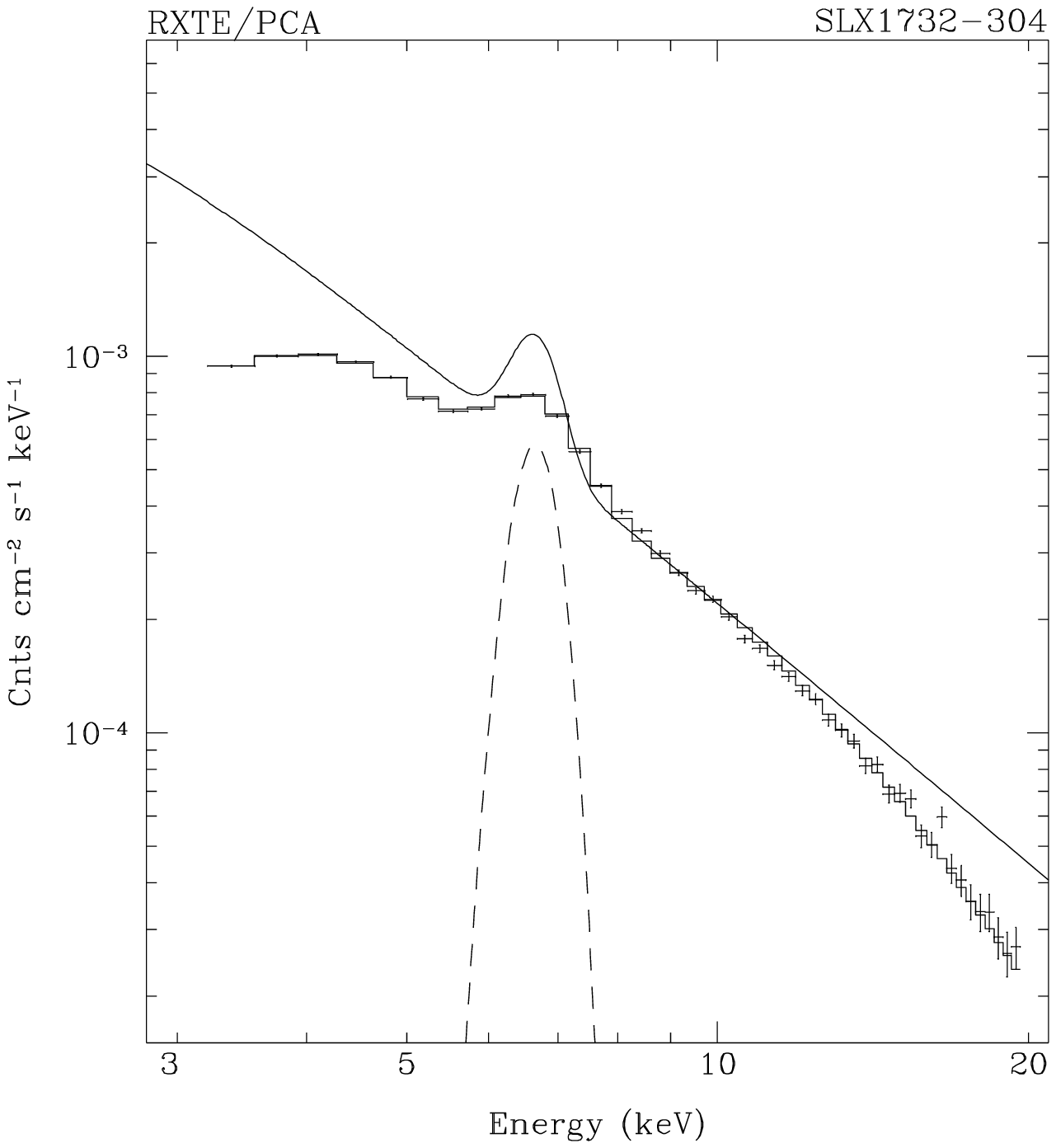}
\vspace{0mm}\caption{RXTE/PCA pulse-height spectrum of SLX 1732-304. 
The histogram and the solid line indicate its best fit and the
corresponding photon spectrum, respectively. 
\label{fig:sp_rxte}}
\end{minipage}
\hspace{9mm}\begin{minipage}{70mm}
\epsfxsize=70mm
\hspace{-5mm}\epsffile[90 295 450 715]{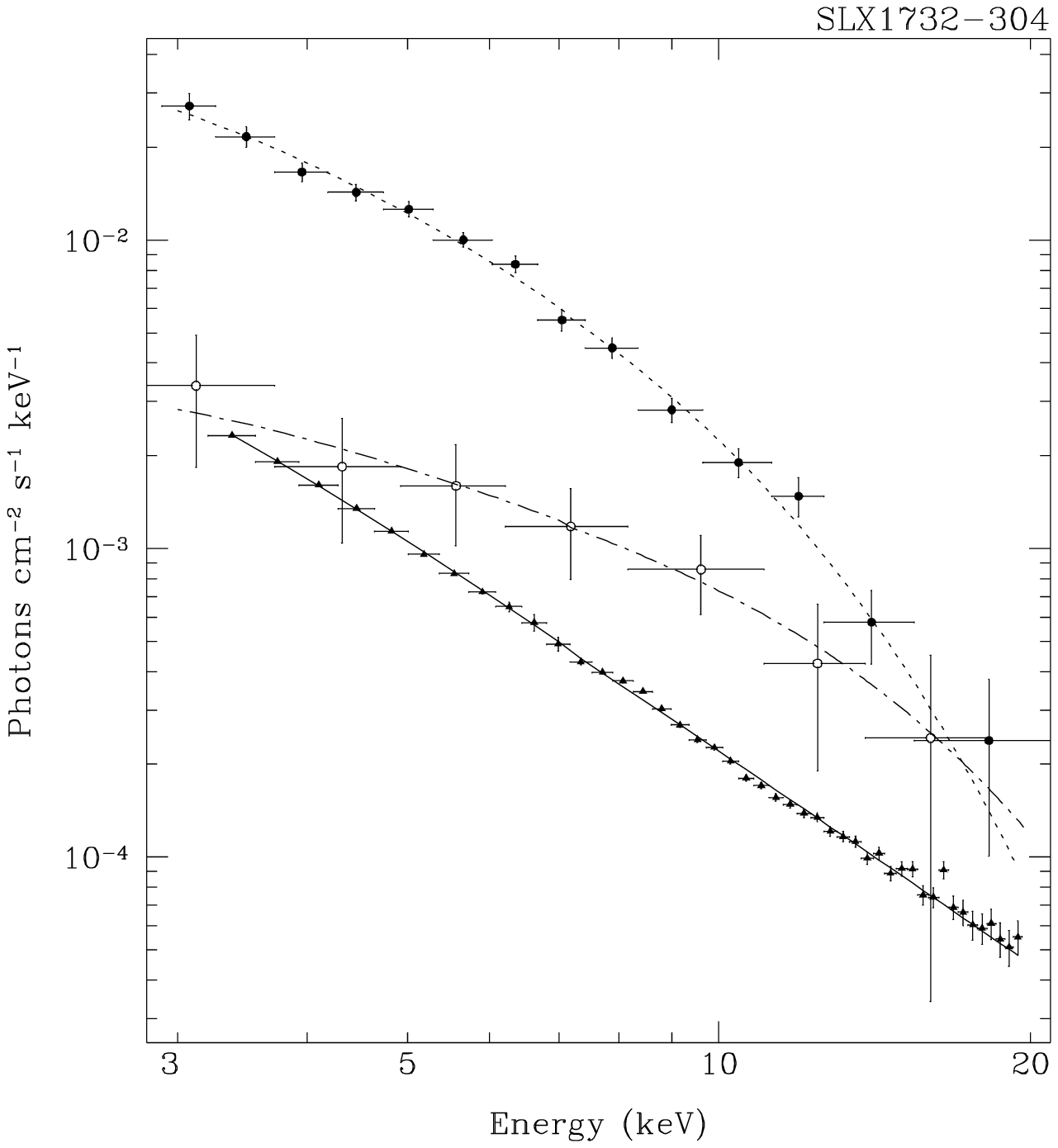}
\vspace{0mm}\caption{Photon spectrum of the persistent X-ray emission 
(triangles) observed by PCA from SLX 1732-304 in comparison with the
source's spectra measured by ART-P during low (open circles) and high
(filled circles) states in 1990.
\label{fig:sp_artp_rx}} 
\end{minipage}
}
\end{figure}

The X-ray photon spectrum of SLX 1732-304 reconstructed from the PCA
data obtained in April 1997 is shown in Fig.\,\ref{fig:sp_rxte}
(Molkov \etal~\cite{mo2}). This spectrum can be good described by
simple power-law model with photon index $\alpha\simeq2.3$. The
measured spectrum exhibits a powerfull emission line at energy
$\sim6.7$ keV with intensity $F_{6.7}\simeq(4.92\pm0.22)\times10^{-4}$
phot.~cm$^{-2}$~s$^{-1}$. This line is most likely unrelated to the
source SLX 1732-304 itself, but is a superposition of the diffuse
6.64-, 6.67-, 6.68-, and 6.7-keV lines of Fe XXV, whose ions recombine
in clouds of hot plasma near the Galactic center.

The X-ray photon spectrum of SLX 1732-304 without the background
diffuse 6.7-keV emission is shown in Fig.\,\ref{fig:sp_artp_rx}
together with source's spectra observed by ART-P on Sept. and
Oct. 1990 during its high and low states (Pavlinsky
\etal~\cite{pa2}). The flux during the former observation $28.6\pm0.7$
mCrab was a factor of $\sim 4$ higher than that during the latter
observation $6.7\pm1.1$~mCrab. The source spectrum in high-state could
be satisfactorily described by bremsstrahlung of an optically thin
thermal plasma with $kT\sim6$ keV or, equally well, by Comptonization
of low-frequency photons in a cloud of hot ($kT_e\sim2.3$ keV)
electron plasma. In the low state, the source most likely had a
power-law spectrum ($\alpha\sim1.7$) with no evidence of an obvious
high-energy cutoff.

The X-ray burst observed with ART-P on September 8 was the third burst
detected from this source over the entire period of its
observations. Our study of the source's spectral evolutions during the
burst indicates that it can be satisfactorily described by the
blackbody model with a temperature smoothly falling during the burst
from 3.8 to 1.0 keV. The total energy release during the burst is
$E\simeq1.7\times10^{39}$ erg, which is equivalent to an explosion of
$9.5\times10^{20}$ g of matter. As was mentioned above we can estimate
the time it takes for this amount of matter to be accreted onto the
stellar surface, i.e., the characteristic burst recurrence time:
$\tau\simeq14^{\rm h}.8$ during the high state and $\tau\simeq62^{\rm
h}.7$ during the low state.

\subsection{4U1724-307}

The time history of the extremely bright X-ray burst which was
detected by the ART-P from this source (Grebenev \etal~\cite{gr}) show
a very interesting feature, the so-called precursor event
(Fig.\,\ref{fig:obzor}). The precursor was observed a few seconds
before the beginning of the primary peak that was a direct evidence
for the nearly Eddington luminosity and quick photospheric radius
expansion of a neutron star in \4u.

\begin{figure}[t]
\hbox{
\begin{minipage}{70mm}
\epsfxsize=70mm
\hspace{-0mm}\psfig{figure=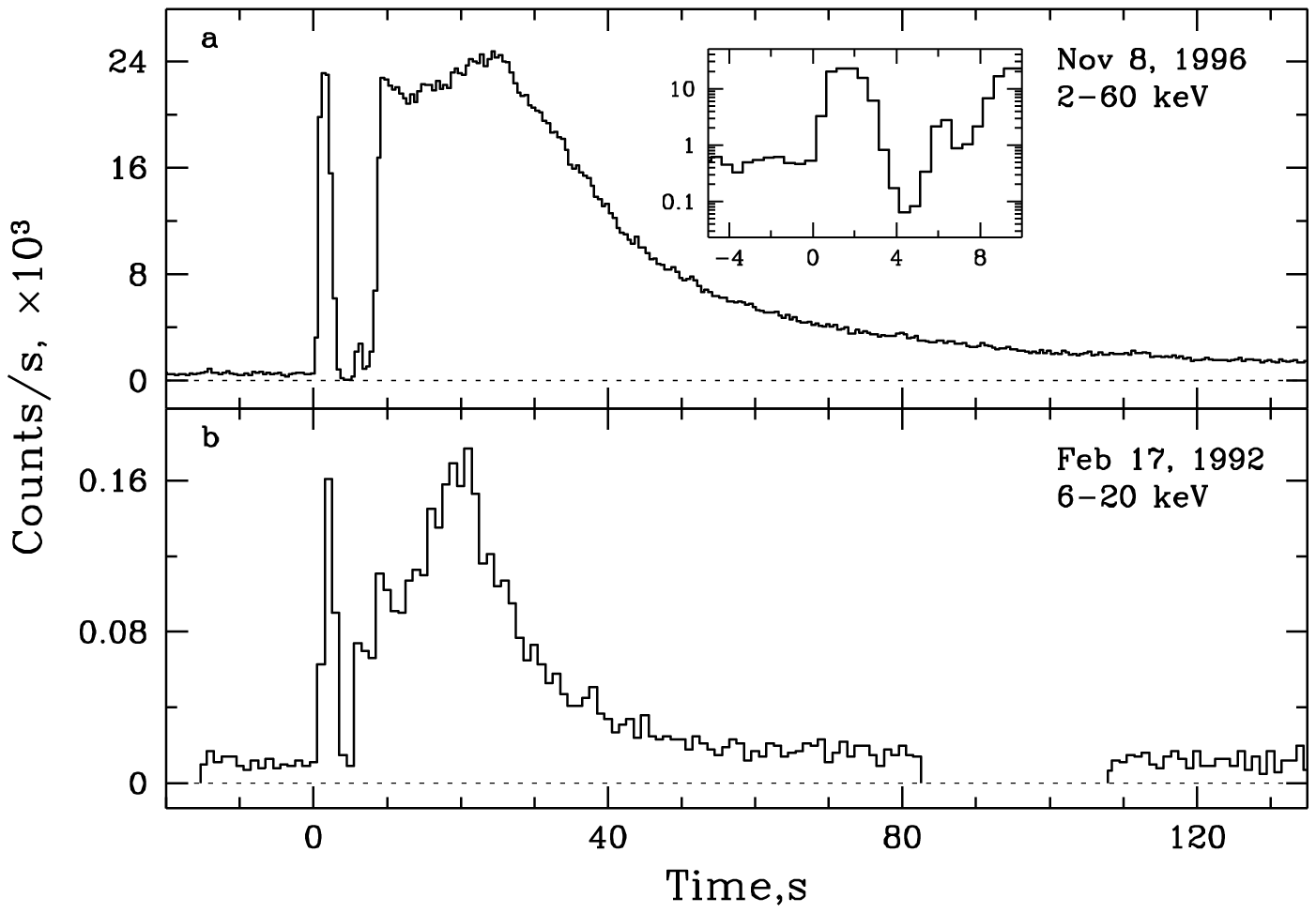,height=1.9in}
\caption{Temporal profiles of the bursts detected by the 
RXTE PCA ({\it a}) and the ART-P ({\it b}) in the broad energy
bands. \label{fig:tim_his}}
\end{minipage}
\hspace{9mm}\begin{minipage}{70mm}
\epsfxsize=70mm
\hspace{-0mm}\psfig{figure=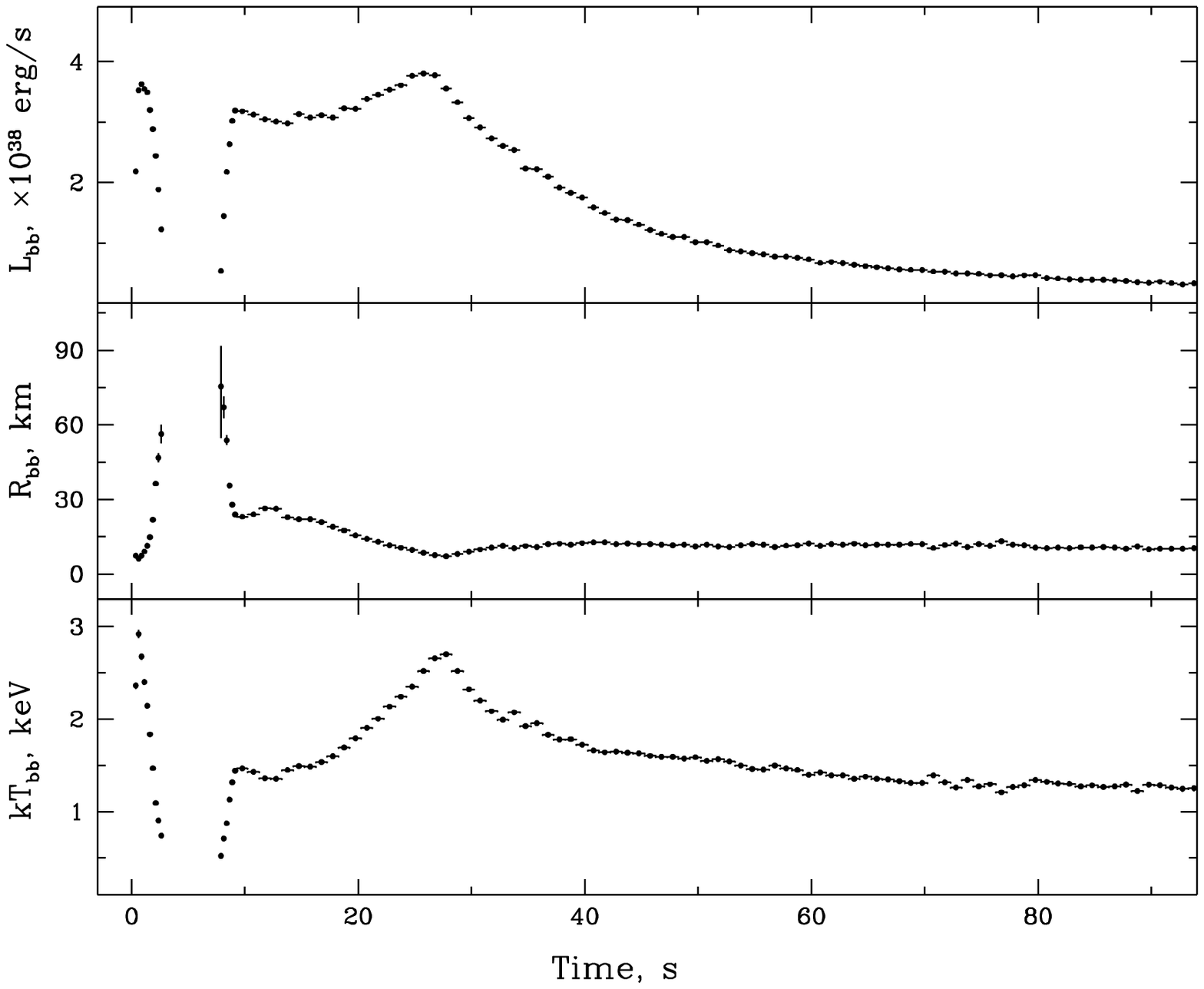,height=2in}
\vspace{0mm}\caption{Evolution of the bolometric luminosity, radius 
and effective temperature of the neutron star photosphere during the
burst in \4u. \label{fig:param}} 
\end{minipage}}
\end{figure}

In 1996-1998 \4u\ was observed by RXTE many times with a total
exposure exceeding 330~ks but with the only X-ray burst detected on
Nov. 8, 1996 (Molkov \etal~\cite{mo3}). In general the observed
profile was very similar to that detected by ART-P
(Fig.\,\ref{fig:tim_his}): the burst also began with an extremely
strong ($\sim 1.9$ Crab) precursor event which lasted $\sim 3$~s, then
a 2-s quiet interval was observed during which the flux fell by
$7.8\pm1.3$ times from the level measured before the burst; the
primary event itself was $\sim 2$ Crab in a peak and had rather a
complex shape and a duration in excess of 150~s.

For spectral analysis we got more than 100 consecutive photon spectra
which were approximated in the \mbox{3-20} keV band with the black
body model. The information on the evolution of spectral parameters
($kT$ and $R$ -- the black body radius of the neutron star
photosphere) and the source bolometric luminosity is presented in
Fig.\,\ref{fig:param}. The figure shows that the initial stage of the
burst was connected with strong photospheric expansion (during the
precursor) and contraction (during the primary event). The observed
radii were in excess of 70 km, velocities -- of $100~
\mbox{km\,s}^{-1}$. While expanding, the photosphere got cool, its
spectrum softened and the fraction of photons with energies within the
PCA band decreased. That was the reason why the first narrow precursor
appeared in the burst profile.

\section*{Acknowledgments}
A.L. and S.M. thanks to Organizing Committee for the providing a
financial support for participation in the Conference.

\end{document}